\documentclass[manuscript]{aastex}
\usepackage{spr-astr-addons}

\shorttitle{Kantowski-Sachs Universe}

\shortauthors{Khadekar et al.}

\begin{document}

\title{Kantowski-Sachs Universe in the Varying Speed of Light Theory}

\author{G.S. Khadekar\altaffilmark{1}}
\altaffiltext{1}{Department of Mathematics, R.T.M. Nagpur
University, Mahatma Jyotiba Phule Educational Campus, Amravati
Road, Nagpur 440033, Maharastra, India\\gkhadekar@yahoo.com}

\author{Arti
Ghogre\altaffilmark{2}} \altaffiltext{2}{Department of
Mathematics, Shri Datta Meghe Polytechnic, Wanadogri, Hingna Road,
Nagpur 441110, India\\arti\_ghogre@rediffmail.com}

\author{Saibal Ray\altaffilmark{3}}
\altaffiltext{4}{Department of Physics, Government College of
Engineering and Ceramic Technology, Kolkata 700010, West Bengal,
India\\saibal@associates.iucaa.in}

\begin{abstract}
In this work we consider the Kantowski-Sachs (KS) universe in the
framework of varying speed of light theory. We present the general
solutions of the gravitational field equations with variable speed
of light $c(t)$, gravitational coupling parameter $G(t)$ and the
decaying vacuum energy $\Lambda(t)$ for the KS model. In the
limiting case for the equation of state (EOS) parameter $\gamma=2$
(stiff fluid with $p=\rho c^2$) and $\gamma =1$ (dust with
$p=0$), exact solutions of the field equations are obtained.
The numerical solutions are also presented for both the
cases. Moreover, it is shown that in the limiting case of large
time, the mean anisotropy parameter tends to zero for $\gamma=2$
and $\gamma=1$. Thus the time variation of the fundamental
constants provides an effective mechanism for the isotropization
of the KS universe.
\end{abstract}


\section{Introduction}

\citet{Dirac1938} proposed the large number hypothesis (LNH),
motivated by the occurrence of large numbers in the universe.
However, in this connection an exclusive review on the LNH can be obtained in the work of~\citet{Ray2007a}
for further reading. Inspired by this theory several
scientists~\citep{Abdel-Rahman1988,Abdel-Rahman1992,Beesham1993,Ray2004,Ray2007b,Ray2008,Mukhopadhyay2008a,Mukhopadhyay2008b,Usmani2008,Ray2009,
Mukhopadhyay2009,Mukhopadhyay2010a,Mukhopadhyay2010b,Ray2011,Hossein2012,Ray2013,Rahaman2013,Mukhopadhyay2016} have been intensively investigated problems on the
variable gravitational constant and cosmological constant with
variable cosmological term $\Lambda$ in the cosmological as
well as astrophysical realm. On the other hand, works have been
done on the cosmological models with variable cosmological term
within a framework of dissipative thermodynamics as well as in the
case of perfect fluid \citep{Vishwakarma2005,Belinchon2006,Belinchon2008}. As $G$ couples geometry to matter, it is
reasonable to consider $G = G(t)$ in an evolving universe when one
considers $\Lambda=\Lambda(t)$. Many extensions of general
relativity with $G = G(t)$ has been made ever since Dirac first
consider the possibility of variable $G$, though none of these
theories has gained wide acceptance \citep{Brans1961,Wesson1978,Wesson1980,
Canuto1977,Weinberg1971,Norman1986,Levit1980,Abdel-Rahman1990}.

The varying speed of light (VSL) cosmologies was proposed by~\citet{Moffat1993}
as an alternatives to cosmological inflation to provide different
basis for resolving the problems of the standard models. He conjectured
that a spontaneous breaking of the local Lorentz invariance and
diffeomorphism invariance associated with a first order
phase-transition can lead to the variation of the speed of light
in the early universe. This idea was later on considerably revived
by~\citet{Albrecht1999} and~\citet{Barrow1999,Barrow2000}. Barrow showed that the
conception of VSL can lead to the solution of flatness, horizon
and monopole problems if the speed of light falls at an
appropriate rate. \citet{Alexander2000} widely studied the
dynamics of VSL in theoretical as well as empirical context.
It has in particular been speculated that VSL offers new paths
of solving the problems of the standard Big Bang cosmology which are
distinct from their resolutions in context of the inflationary
paradigm~\citep{Guth1981} or the pre-Big Bang scenario of low
energy string theory~\citep{Veneziano1997}. Moreover, in contrast
to the case of inflationary Universe, VSL may provide an
explanation for the relativistic smallness of the various
physical constants today.

In connection to VSL theory~\citet{Barrow1999} introduced a Machian
scenario in which $c=c_{0}a^{n}$, where $a$
is the scale factor. This has significant advantages to the
phase-transition scenario in which the speed of light changes
suddenly from $c^{-}$ to $c$. On the other hand,~\citet{Albrecht1999}
have investigated possible consequences of a time variation in
the velocity of light in vacuum. The Einstein field equations
for FRW spacetime in the VSL theory have been solved by~\citet{Barrow1999,Barrow2000}
for anisotropic model, who also obtained the rate of variation of
the speed of light required to solve flatness and cosmological constant
problems. Some other mentionalable works in this field are as follows:
(i) By assuming energy conservation of observed
matter \citet{Gopakumar2001} have solved the flatness
cosmological constant problem with varying speed of light $c$,
gravitational coupling strength $G$ and cosmological parameter
$\Lambda$, (ii)~\citet{Chimento2001} have found exact constant solutions
for cosmological density parameter using generalization of general
relativity that incorporates a cosmic time variation of velocity
of light in vacuum and Newtonian gravitation constant, and (iii)~\citet{Belinchon2005,Belinchon2006,Belinchon2007} studied perfect
fluid Bianchi type I model with variable $G$, $c$ and $\Lambda$.

Besides all the above mentioned major works on VSL theory it is
observed that some authors~\citep{Harko1999,Shojaie2006} have proposed
a new generalization of general relativity which also allows arbitrary
changes in the speed of light $c$ and the gravitational constant
$G$. However, this has been done in such a way that variation in
the speed of light introduces corrections to the curvature tensor
in the Einstein equations in the cosmological frame.  \citet{Harko1999}
considered the evolution and dynamics of Bianchi type I and V Universe
and obtained exact solutions of the gravitational field equations
in a small time scale limit.

The purpose of the present paper is to extend the results previously
obtained in the framework of the homogeneous and isotropic FRW
cosmological models to the case of anisotropic KS universe. In essence
we carried out here an investigation to highlight specific features of
KS Universe which basically represents dust solutions to the Einstein
field equation and are a widely used family of inhomogeneous cosmological
models. Basically, we have generalized the work of~\citet{Harko1999}
by considering that the constants are functions of the volume scale factor
and obtained the solutions of the field equations under the framework
of VSL theory in the small time limit. The paper is organized as follows:
The field equations for KS model are written down in Sec. 2. In Sec. 3
we have obtained exact solutions of the field equations for VSL model
corresponding to specific time variation law of constants. In Sec. 4,
we conclude our results.

\section{The Einstein field equations}

We consider the KS Universe in the framework of Einstein's general
relativity. As an additional condition we impose on the physical
constants some restrictions as advocated by the LNH and VSL
theory. The line element for KS Universe is given by
\begin{equation}
\label{eq1}  ds^{2}= -c^{2}(t)dt^{2}+a_{1}^{2}(t)dr^{2}+a_{2}^{2}(d\theta^{2} + sin^{2}\theta d\phi^{2}),
\end{equation}
where  $a_{1}$, $a_{2}$ are the scale factors.

The Einstein field equations take the usual form
\begin{equation}
\label{eq2} R_{ij}-\frac{1}{2}g_{ij}R = - \frac{8 \pi
G(t)}{c^{4}(t)}T_{ij}+\Lambda(t)g_{ij}.
\end{equation}

The energy momentum tensor can be written as
\begin{equation}
\label{eq3} T_{ij} = (p+\rho \ c^{2}) u_{i} u_{j} + p g_{ij},
\end{equation}
where $u^{i}$~($i = 0,1,2,3$) be the four velocity, $p$ and $\rho$ are respectively the fluid pressure and energy density.

For the KS metric (1), the Einstein field Eq. (2) can be written as
\begin{equation}
\label{eq4}\frac{\dot{a_{2}}^{2}}{a_{2}^{2}}+\frac{2\dot{a_{1}}\dot{a_{2}}}{a_{1}a_{2}}+\frac{c^{2}}{a_{2}^{2}}=8 \pi G\rho+\Lambda c^{2} ,
\end{equation}

\begin{equation}
\label{eq5}\frac{2\ddot{a_{2}}}{a_{2}}+\frac{\dot{a_{2}}^{2}}{a_{2}^{2}}-\frac{2\dot{a_{2}}\dot{c}}{a_{2}c}+\frac{c^{2}}{a_{2}^{2}}=
- \frac{8 \pi Gp}{c^{2}}+\Lambda c^{2},
\end{equation}

\begin{equation}
\label{eq6}\frac{\ddot{a_{1}}}{a_{1}}+\frac{\ddot{a_{2}}}{a_{2}}+\frac{\dot{a_{1}}\dot{a_{2}}}{a_{1}a_{2}}
-\frac{\dot{a_{2}}\dot{c}}{a_{2}c}-\frac{\dot{a_{1}}\dot{c}}{a_{1}c}=
- \frac{8 \pi Gp}{c^{2}}+\Lambda c^{2}.
\end{equation}
We assume that the thermodynamic pressure $p$ of the cosmological fluid obeys a linear equation of state
\begin{equation}
\label{eq7}p=(\gamma - 1)\rho c^{2}(t),
\end{equation}
where equation of state parameter $\gamma =$ constant and $1 \leq
\gamma \leq 2$.

The conservation equation gives
\begin{equation}
\label{eq8}\dot{\rho}+ \left(\frac{\dot{a_{1}}}{a_{1}}+
\frac{\dot{2a_{2}}}{a_{2}} \right)\gamma \rho =
-\frac{c^{2}\dot{\Lambda}}{8 \pi G}-\frac{\dot{G}}{G} \rho +
\frac{2\dot{c}}{c}\rho.
\end{equation}

We assume that $\left(T_{i;j}^{j}\right) = 0$, which leads to
the following two equations:
\begin{equation}
\label{eq9}\dot{\rho}+ \left(\frac{\dot{a_{1}}}{a_{1}}+ \frac{\dot{2a_{2}}}{a_{2}} \right)\gamma \rho = 0,
\end{equation}

\begin{equation}
\label{eq10} -\frac{c^{2}\dot{\Lambda}}{8 \pi G}-\frac{\dot{G}}{G}\rho + \frac{2\dot{c}}{c}\rho = 0.
\end{equation}

For later convenience we introduce the following variables:
\begin{equation}
\label{eq11}V = a_{1} a_{2}^{2},
\end{equation}

\begin{equation}
\label{eq12}H_{i}=\frac{\dot{a_{i}}}{a_{i}},
\end{equation}

\begin{equation}
\label{eq13}H = \frac{1}{3}(H_{1}+2 H_{2}).
\end{equation}

We further define
\begin{equation}
\label{eq14}\Delta H_{i} = H_{i} - H,
\end{equation}
where $i = 1, 2, 3$ and $H_{2} = H_{3}=
\frac{\dot{a_{2}}}{a_{2}}$, the factors $V$, $H_{i}$ and $H$ being
the volume scale factor, directional Hubble parameters and mean
Hubble parameter respectively.

We obtain
\begin{equation}
\label{eq15}H = \frac{\dot{V}}{3V}.
\end{equation}

In addition to this we introduce the physical quantities related
to cosmology as follows:
\begin{equation}
\label{eq16}\theta = 3 H,
\end{equation}

\begin{equation}
\label{eq17}A=\frac{1}{3}\sum_{i=1}^{3}\left(\frac{\Delta
H_{i}}{H}  \right)^{2}=
\frac{1}{3}\sum_{i=1}^{3}\frac{(H_{i}-H)^{2}}{H^{2}},
\end{equation}

\begin{equation}
\label{eq18}\sigma^{2} = \frac{1}{2}\sigma_{ik} \sigma^{ik} =
\frac{1}{2}\left(\sum_{i=1}^{3}H_{i}^{2}-3H^{2}\right)=\frac{3AH^{2}}{2},
\end{equation}

\begin{equation}
\label{eq19}q = -H^{-2}(\dot{H}+H^{2}) =
\frac{d}{dt}\left(\frac{1}{H} \right)-1,
\end{equation}
where $\theta$, $A$, $\sigma^{2}$ and $q$ are the scalar
expansion, mean anisotropy parameter, shear scalar and
deceleration parameter respectively.

Note that $A = 0$ for isotropic expansion. Moreover, the signature
of the deceleration parameter indicates how the Universe expands.
In fact the positive sign corresponds to `standard' deceleration
model whereas a negative sign indicates an accelerating Universe.

\section{An exact solution for the KS universe}

In this section we find the exact solutions of the field equations.\\ From Eq. (5) and (6), we get
\begin{equation}
\label{eq20}\frac{\dot{a_{2}}^{2}}{a_{2}^{2}}+\frac{\ddot{a_{2}}}{a_{2}}+\frac{c^{2}}{a_{2}^{2}}
-\frac{\dot{a_{2}}\dot{c}}{a_{2}c}+\frac{\dot{a_{1}}\dot{c}}{a_{1}c}=\frac{\dot{a_{1}}\dot{a_{2}}}{a_{1}a_{2}}+\frac{\ddot{a_{1}}}{a_{1}}.
\end{equation}

Adding Eq. (4) and (5) and using Eq. (20) in it, we obtain
\begin{equation}
\label{eq21}\frac{1}{V}\frac{d}{dt}(VH_{1})-\frac{\dot{c}}{c}H_{1}=4\pi \ G \left(\rho-\frac{p}{c^{2}}\right)+\Lambda c^{2}.
\end{equation}

Also adding Eq. (4) and (6) and using Eq. (20), we get
\begin{equation}
\label{eq22}\frac{1}{V}\frac{d}{dt}(VH_{2})-\frac{\dot{c}}{c}H_{2}=4\pi \ G \left(\rho-\frac{p}{c^{2}}\right)+\Lambda c^{2}-\frac{c^{2}a_{1}}{V}.
\end{equation}

From Eqs. (21) and (22) with the help of Eq. (12) one can get
\begin{equation}
\label{eq23} 3\dot H + H_{1}^2 + 2H_{2}^2-3  \frac{\dot c}{c}H= -4 \pi \ G \left(\rho+3\frac{p}{c^{2}}\right)+\Lambda c^{2}.
\end{equation}

Again adding Eq. (21) and (22), we get
\begin{eqnarray}
\label{eq24}\frac{1}{V}\frac{d}{dt}(V[H_{1}+2H_{2}])-\frac{\dot{c}}{c}(H_{1}+2H_{2})\nonumber
\\  =12\pi \ G \left(\rho-\frac{p}{c^{2}}\right)+3\Lambda
c^{2}-\frac{2c^{2}a_{1}}{V}.
\end{eqnarray}

Now using Eq. (13) and (15) in Eq. (24), with (7) we get
\begin{equation}
\label{eq25}\ddot{V}-\dot{V}\frac{\dot{c}}{c}=3\Lambda V c^{2}+12\pi \ G(2-\gamma)\rho V - 2 a_{1}^{2}c^{2}.
\end{equation}

Now we assume that the `constants' $G$, $c$ and $\Lambda$ are
decreasing functions of time and to describe their variations we
use the following simple phenomenological law given
by~\citep{Harko1999}
\begin{equation}
\label{eq26}G = G_{0}+\frac{G_{1}}{V^{\alpha}}, \;\;\;   \Lambda=\Lambda_{0}+\frac{\Lambda_{1}}{V^{\beta}}, \ \  \ \  c=c_{0}+\frac{c_{1}}{V^{\eta}}.
\end{equation}
where $G_{0}>0, G_{1}\ge 0, \Lambda_{0}>0, \Lambda_{1}\ge0, c_{0}>0, c_{1}>0, \alpha >0, \beta>0$ and $\eta>0$ are all constants.\\

For $t \rightarrow 0$, $V$ is extremely small, then
\begin{equation}
\label{eq27} G \approx \frac{G_{1}}{V^{\alpha}}, ~ \Lambda \approx
\frac{\Lambda_{1}}{V^{\beta}}, ~ c \approx \frac{c_{1}}{V^{\eta}}.
\end{equation}

From the conservation Eq. (9), $\rho$ can be expressed as
\begin{equation}
\label{eq28}\rho = \rho_{0}V^{-\gamma},
\end{equation}
where $\rho_{0}$ is constant of integration.

After using Eq. (27) and (28) in Eq. (10), we get the consistency
conditions relating the constants $\alpha, \beta, \gamma, \eta,
G_{1}, \Lambda_{1}$ and $c_{1}$ as
\begin{equation}
\label{eq29} \alpha -\beta + \gamma - 2\eta = 0,
\end{equation}

\begin{equation}
\label{eq30}\alpha - 2\eta = \frac{-c_{1}^{2} \Lambda_{1} \beta}{8 \pi G_{1}\rho_{0}}.
\end{equation}

Hence Eq. (29) and (30), yield
\begin{equation}
\label{eq31}\beta = \left(1+ \frac{c_{1}^{2}\Lambda_{1}}{8 \pi G_{1}\rho_{0}} \right)^{-1}.
\end{equation}

Now by using Eq. (27) and (28) in Eq. (25), we get
\begin{equation}
\label{eq32}V\ddot{V}+ \eta \dot{V}^{2}=\frac{3c_{1}^{2}\Lambda_{1}}{V^{2\eta +\beta - 2}} +\frac{12 \pi G_{1}\rho_{0}(2-\gamma)}{V^{\alpha + \gamma-2}}-\frac{2a_{1}c_ {1}^{2}}{V^{2 \eta}-1}
\end{equation}

From Eq. (32), after first integration we get
\begin{equation}
\label{eq33} \dot{V}^{2}=V^{-2 \eta}F(V),
\end{equation}
where
\begin{equation}
\label{eq34}F(V)=\left(\frac{6c_{1}^{2}\Lambda_{1}}{2-\beta}V^{2-\beta}+\frac{24 \pi G_{1}\rho_{0}(2-\gamma)}{2 \eta-\alpha-\gamma +2}V^{(2\eta-\alpha-\gamma+2)}+C \right).
\end{equation}
where  $C= -4c_{1}^{2}\int a_{1}dV +c_{2}$ and $c_{2}>0$ is a constant of integration.

From  Eqs. (21) and  (22) after  integration, we get
\begin{equation}
\label{eq35} log[V(H_{1}-H_{2})]=log c + log K,
\end{equation}
where $ K =exp\left[\int \frac{c^{2}a_{1}}{V(H_{1}-H_{2})}dt
\right]$.

We note that~\citet{Obukhov2006} considered $K=\int a_{1}dt$ in the conventional Einstein's theory.
In our case the value of $K$ is similar to this value for $\alpha =\beta=\eta = 0$ and $c_{1}=1$.

After simplification Eq. (35) provides
\begin{equation}
\label{eq36} H_{1}=H+\frac{2}{3} \frac{Kc}{V},
\end{equation}
and
\begin{equation}
\label{eq37} H_{2}=H - \frac{Kc}{3V}.
\end{equation}

If we substitute Eqs. (36) and (37) in Eq. (23), the with the
help of Eqs. (29) and (30) we get
\begin{eqnarray}
\label{eq38} \frac{c_{1}^2\Lambda_{1}\beta[\gamma (\eta
+\beta)-2\eta +2]}{(\beta -1)(\beta-2) V^{(2\eta +\beta-2)}}
-\frac{2}{V^{2\eta}}\left(a_{1}c_{1}^2 V +\frac{c_{2}}{3}\right) \nonumber \\ +
\frac{8 c_{1}^2}{3V^{2\eta}}\int a_{1} dV +\frac{2}{3}K^2=0.
\end{eqnarray}

For $\alpha=\beta= \eta=c_{2}=0$ and  $ c_{1}=1$, the above
equation reduces to the equation of the form $$ -2a_{1}V +
\frac{8}{3}\int a_{1} dV +\frac{2}{3}\left( \int a_{1}
dt\right)^2=0,$$ as obtained earlier by~\citet{Obukhov2006} in the
general theory of relativity.

By taking $V\geq 0$, as a parameter, we can obtain the general
solution for KS model with physical quantities as follows:
\begin{equation}
\label{eq39} t-t_{0}=\int \frac{V^{\eta}}{F(V)^{1/2}}dV,
\end{equation}

\begin{equation}
\label{eq40}\theta=3H =\frac{F(V)^{1/2}}{V^{\eta+1}},
\end{equation}

\begin{equation}
\label{eq41}
a_{1}=V^{1/3}a_{01} \  exp\left[\frac{2}{3}Kc_{1}\int \frac{dV}{V F(V)^{1/2}}\right],
\end{equation}

\begin{equation}
\label{eq42}a_{2}=V^{1/3}a_{02} \  exp\left[\frac{-1}{3}Kc_{1}\int \frac{dV}{V F(V)^{1/2}}\right],
\end{equation}

\begin{equation}
\label{eq43}A= \frac{2K^{2}c_{1}^{2}}{F(V)},
\end{equation}

\begin{equation}
\label{eq44}\sigma^{2}=\frac{K_{1}^{2}c_{1}^{2}}{3V^{2(\eta +1)}}.
\end{equation}

\begin{eqnarray}
\label{eq45}q = 3\eta +2+\frac{3V}{2F(V)} \times \nonumber \\
\left[-6\Lambda_{1}c_{1}^{2}V^{1-\beta} +24 \pi
G_{1}\rho_{0}(\gamma - 2)V^{(2\eta-\alpha - \gamma+1)} +
4c_{1}^{2}a_{1}\right],
\end{eqnarray}
where $t_{0},\; a_{01}$ and $ a_{02}$ are constants of
integration.

\subsection{Case I : $\gamma = 2$}

In this case for extremely small  $V$,  Eq. (34) can be written as
\begin{equation}
\label{eq46}F(V)=\frac{6c_{1}^{2}\Lambda_{1}}{2-\beta}V^{2- \beta}+ B_{1},
\end{equation}
where $B_{1}$ is constant.

Using Eq. (46) in Eq. (33) we get
\begin{equation}
\label{eq47} \dot{V}^{2}=V^{-2 \eta}\left(\alpha_{0}V^{2-\beta}+B_{1}\right),
\end{equation}
where $\alpha_{0}=\left(\frac{6c_{1}^{2}\Lambda_{1}}{2-\beta}\right),\ \  \beta \neq 2$.

After integrating Eq. (47) with  the condition
$\alpha_{0}V^{2-\beta} >> B_{1}$  and  $t_{0}=0$, it gives
\begin{equation}
\label{eq48}V \approx t^{\frac{2}{2\eta+\beta}}\Longrightarrow H \approx \frac{1}{t}.
\end{equation}

Hence the scale factors $a_{1}$ and $a_{2}$ can be obtained as
\begin{equation}
\label{eq49}a_{i} \approx t^{\frac{2}{3(2\eta + \beta)}},
\end{equation}
for $i = 1,2$.

Using Eqs. (43) and (48) we get
\begin{equation}
\label{eq50}A \approx t^{\frac{2(\beta -2)}{(2 \eta+\beta)}}, \;\; \beta < 2.
\end{equation}
\begin{equation}
\label{eq51}\sigma^{2} \approx  \left(\frac{1}{t^{\frac{4(\eta +1)}{2 \eta +\beta}}}\right).
\end{equation}
\begin{equation}
\label{eq52}q = 3\left(\eta +\frac{\beta}{2}\right)-1.
\end{equation}

Thus $q$ tends to a constant value. The analogous solutions for
extremely small value of $V$ are discussed by~\citet{Harko1999}
for Bianchi type I model.

\subsection{Case II : $\gamma = 1 $}
Using Eq. (34) with $ \Lambda_{1}=0$ for extremely small value of
$V$, we get
\begin{equation}
\label{eq53}F(V)=b_{0}V^{2\eta-\alpha+1}+B_{3},
\end{equation}
where $b_{0}=\left(\frac{24\pi G_{1}\rho_{0}}{(2\eta -\alpha
+1)}\right)$ and $B_{3}$= constant.

Hence using Eq. (53) in Eq. (33), we get
\begin{equation}
\label{eq54} \dot{V}^{2}=V^{-2 \eta} \left(b_{0}V^{(2\eta-\alpha+1)}+B_{3}\right).
\end{equation}

After integrating Eq. (54) with the condition
$b_{0}V^{2\eta-\alpha +1} >> B_{3}$  and  $t_{0}=0$, it gives
\begin{equation}
\label{eq55}V \approx t^{\frac{2}{3-\alpha}}\Longrightarrow H \approx \frac{1}{t}.
\end{equation}

The other variables can be calculated straightforwardly as
follows:
\begin{equation}
\label{eq56}a_{i} \approx  t^{\frac{2}{3(3-\alpha)}},\;\alpha \neq 3,
\end{equation}
for $i = 1, 2$.

\begin{equation}
\label{eq57}A\approx t^{\frac{-2(2\eta +1-\alpha)}{(3-\alpha)}},\;\alpha \neq 3.
\end{equation}
\begin{equation}
\label{eq58}\sigma^{2} \approx  t^{\frac{-4(\eta +1)}{(3-\alpha)}},\; \alpha \neq 3.
\end{equation}
\begin{equation}
\label{eq59}q = \frac{3}{2}(\alpha-1)+2.
\end{equation}
 The deceleration parameter is always positive when  $\alpha > -1/3$ and negative when  $\alpha <-1/3$.

\begin{figure}[h]
\begin{center}
\includegraphics[width=0.4\textwidth]{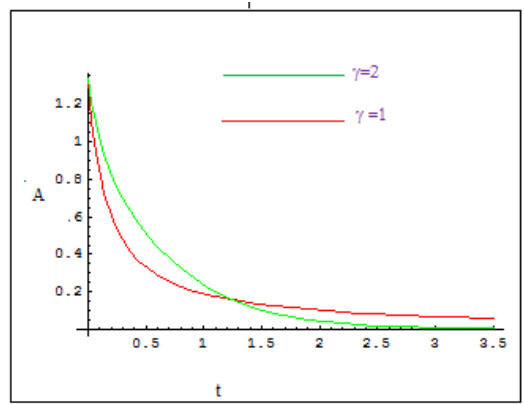}
\caption{Variation with respect to time of the mean anisotropy
parameter  for KS model with perfect cosmological fluid for
$\gamma = 2$ (green line) and $\gamma = 1$ (red line), in the
small time limit. The constants are chosen as: $\beta=1, c_{1}=1,
C=1,\Lambda_{1}=1 $}
\end{center}
\end{figure}

\begin{figure}[h]
\begin{center}
\includegraphics[width=0.4\textwidth]{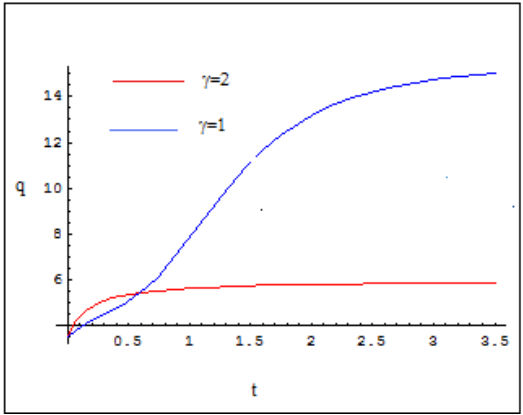}
\caption{ Variation with respect to the time of the deceleration
parameter $q$ for stiff fluid $\gamma = 2$ (Red line), $\gamma =
1$ (blue line) and  in the small $V$ limit. The constants are
chosen as: $\beta=1,\eta = 1/2, c_{1}=1, C=1,\Lambda_{1}=1, 8\pi
G_{1}\rho_{0} =1, 4c_{1}^{2}a_{1}=1$.}
\end{center}
\end{figure}

\section{Conclusion}
In this paper we have generalized the work of~\citet{Harko1999}
for KS model in the framework of VSL theory. We have obtained the
solutions for KS model in a small time limit in which variables
$G$, $c$ and $\Lambda$ are functions of the volume scale factor.
For extremely small value of $V$ and $\gamma = 2$, with the
condition $\alpha_{0}V^{2-\beta} >> B_{1}$, $ F(V) \propto V^{2-
\beta}$ with $t_{0}=0$, it is observed that $V \propto
t^{\frac{2}{2\eta + \beta}} $ and hence, the expansion of the
early Universe is of the form of power law of the expansion. The
mean anisotropy will increase for $\beta > 2$, and KS spacetime
will not end in isotropic state in large time limit. However, for
$\beta < 2$, the mean anisotropy tends to 0 in the large time
limits, thus the KS type VSL cosmology is providing an effective
mechanism for the isotropization of the Universe. The evolution of
the anisotropic flat Universe is generally non-inflationary, with
the deceleration parameter $q > 0$, and tending, for large
times, to a constant value given in the Eq. (52). The deceleration
parameter is given by $q = 3(\eta +\beta /2)-1 $,  which is always
positive when $\beta > (2/3-2\eta) $ and negative when $\beta < (2/3-2\eta) $.

Similarly, for the case $\gamma = 1$  with  $ \Lambda_{1}=0$ we
have shown that $V \propto t^{\frac{2}{3-\alpha}}$ as $t
\rightarrow 0$. Hence again the expansion of the early Universe is
of the form of power law expansion. For $\alpha > 3$, the mean
anisotropy will increase for large times limit and KS space time
will not end in an isotropic state, whereas for $\alpha < 3$, the
mean anisotropy tends to $0$, thus the KS type VSL cosmology
becomes isotropic. Moreover, the deceleration parameter is given
by $q= \frac{3}{2}(\alpha -1)+2$, which is always positive for
$\alpha > 1$.

The time variation of  mean anisotropic parameter $A$ and
deceleration parameter $q$ for KS spacetime are plotted with
respect to the time as shown in the Figs. 1 and 2.

It have argued by~\citet{Damour1988} that the existence of
cosmological constant $\Lambda$ or variable $G$
is not in conflict with observational determination of the age of
the Universe or with some astrophysical data.
The dynamics and evolution of the Universe is essentially
determined by the values of the constants $\alpha$, $\beta$, and
$\eta$ describing the time variations of $G$, $c$ and $\Lambda$
and which are arbitrary in this model.

However, we would like to conclude here with a final comment that~\citet{Ghosh2012}
investigated the possible variation of $c$ in the context of the
present accelerating Universe as discovered through SN Ia
observations and show that variability of $c$ is not permitted under
the variable $\Lambda$ models. This obviously impose an observational
constraint on the theoretical speculation of VSL theory, however
at the same time allows variation of $\Lambda$ in terms of {\it dark energy}
via the background platform of LNH as put forward by\citet{Dirac1938}.

\section*{Acknowledgement}
The authors are thankful to Prof. Sriram for useful discussions.
This work is partially supported by University Grant Commission,
New Delhi under the Special Assistance Programme No. F.
510/1/DRS/2011(SAP-I). The authors (GSK and AG) thank for
the Inter University Center for Astronomy and Astrophysics
(IUCAA), Pune for providing the necessary literature during visit
under the SAP programme. Also SR is thankful to the authority of
Inter-University Center for Astronomy and Astrophysics, Pune, India
as well as The Institute of Mathematical Sciences, Chennai, India
for providing Associateship under which a part of this work was carried out.

\end{document}